\documentclass[twocolumn,showpacs,preprintnumbers,amsmath,amssymb]{revtex4}
\usepackage{bm,dcolumn,graphicx}
\newcommand{\as}{\alpha_s}
\newcommand{\wh}{\widehat}
\newcommand{\nn}{\nonumber}

\newcommand{\eqn}[1]{(\ref{#1})}
\newcommand{\DKP}{\Delta_{K\pi}}

\newcommand{\mev}{\mbox{\rm MeV}}
\newcommand{\gev}{\mbox{\rm GeV}}
\newcommand{\tvs}{\vbox{\vskip 6mm}}
\newcommand{\sfrac}[2]{\mbox{$\frac{#1}{#2}$}}

\begin{document}

\preprint{UAB-FT-600}
\preprint{IFIC/06-14}
\preprint{FTUV/06-0509}

\title{Scalar \boldmath{$K\pi$} form factor and light quark masses}

\author{Matthias Jamin}
\email{jamin@ifae.es}
\author{Jos\'e Antonio Oller}
\email{Oller@um.es}
\author{Antonio Pich}
\email{Antonio.Pich@uv.es}

\affiliation{${}^*$Instituci\'o Catalana de Recerca i Estudis Avan\c{c}ats
(ICREA), Theoretical Physics Group, IFAE, Universitat Aut\`onoma de Barcelona,
E-08193 Bellaterra, Barcelona,
Spain,}
\affiliation{${}^\dagger$Departamento de F\'{\i}sica, Universidad de Murcia,
E-30071 Murcia, Spain,}
\affiliation{${}^\ddagger$Departament de F\'{\i}sica Te\`orica, IFIC,
Universitat de Val\`encia -- CSIC, Apt. Correus 22085, E-46071 Val\`encia,
Spain.}

\date{August 22, 2006}

\begin{abstract}
Recent experimental improvements on $K$-decay data allow for a precise
extraction of the strangeness-changing scalar $K\pi$ form factor and the
related strange scalar spectral function. On the basis of this scalar as well
as the corresponding pseudoscalar spectral function, the strange quark mass
is determined to be $m_s(2\,\gev)=92\pm 9\,\mev$. Further taking into
account chiral perturbation theory mass ratios, the light up and down quark
masses turn out to be $m_u(2\,\gev)=2.7\pm 0.4\,\mev$ as well as
$m_d(2\,\gev)=4.8\pm 0.5\,\mev$. As a by-product, we also find a value for the
Cabibbo angle $|V_{us}|=0.2236(29)$ and the ratio of meson decay constants
$F_K/F_\pi=1.203(16)$. Performing a global average of the strange mass by
including extractions from other channels as well as lattice QCD results
yields $m_s(2\,\gev)=94\pm 6\,\mev$.

\end{abstract}


\maketitle


\section{Introduction}

Together with the strong coupling, quark masses are fundamental QCD parameters
of the Standard Model (SM). Precise values are required as an input in many
phenomenological applications and they can also prove useful in constraining
unified theories which go beyond the SM. In the light quark sector, of special
interest is the mass of the strange quark, as it plays an important role in
theoretical predictions of the parameter for direct CP violation
$\varepsilon'/\varepsilon$ and SU(3) breaking in hadronic $B$ decays.

Present day approaches to determine quark mass values include on a more
fundamental level lattice gauge theory, whereas on a more phenomenological
level, operator product expansion (OPE) as well as analyticity properties can
be exploited to relate the quark-gluon picture of QCD to hadronic observables
\cite{svz79}, thereby also allowing for an extraction of quark masses.
Further information on the light quark masses can be obtained from chiral
perturbation theory ($\chi$PT) \cite{gl85}, which permits to make predictions
for quark mass ratios. In this work, the latter two approaches will be
considered to first determine the mass of the strange quark $m_s$ and from
that also the up and down quark masses $m_u$ and $m_d$.

Good sensitivity to $m_s$ is achieved in the analysis of scalar and
pseudoscalar hadronic channels, since the corresponding two-point correlation
functions turn out to be proportional to $m_s^2$. Both channels have already
been studied extensively in the literature. In the scalar channel, after the
pioneering work of \cite{nprt83}, recent analyses have been performed in
refs.~\cite{jop01b,mal99,jam97,cfnp97,bgm97,cps96,cdps94,jm94}, whereas the
pseudoscalar channel has been investigated in refs.~\cite{ck05,mk01}. A
fundamental ingredient for the scalar channel analysis is the scalar strange
spectral function, which has been obtained in a series of articles
\cite{jop00,jop01,jop01b} on the basis of dispersion relations and S-wave
$K\pi$ scattering data.

Recent experimental improvements on $K$-decay data, to be discussed in the
next section, allow for a precise determination of the ratio
$F_K/F_\pi/F_+^{K\pi}(0)$, and consequently for a substantial improvement of
the scalar $K\pi$ form factor and the related strange spectral function which
will be performed in section~3.  In section~4, the scalar strange spectral
function is employed to determine the strange mass $m_s$. For comparison, we
also consider the $m_s$ determination from the pseudoscalar channel in
section~5. Finally, in the conclusions, our final average for $m_s$ will be
presented and compared to recent extractions from lattice QCD, and from our
result for $m_s$ and $\chi$PT mass ratios, also $m_u$ and $m_d$ are deduced.

\section{The ratio \boldmath{$F_K/F_\pi/F_+^{K\pi}(0)$}}

Todays knowledge of the pseudoscalar meson decay constants dominantly 
originates from the leptonic decays $P\to l\nu$ \cite{pdg04}. The ratio of the
decay rates $\Gamma[K\to l\nu_l(\gamma)]$ and $\Gamma[\pi\to l\nu_l(\gamma)]$
is directly proportional to the square of $F_K/F_\pi$. Including electromagnetic
radiative corrections according to Marciano and Sirlin \cite{ms93}, it reads
\begin{eqnarray}
\label{RPl2}
\frac{\Gamma[K\to l\nu_l(\gamma)]}{\Gamma[\pi\to l\nu_l(\gamma)]} &\!\!=\!\!&
\frac{|V_{us}|^2}{|V_{ud}|^2}\frac{F_K^2 M_K}{F_\pi^2 M_\pi}\frac{(1-x_K^2)^2}
{(1-x_\pi^2)^2}\frac{[1+\sfrac{\alpha}{\pi}F(x_K)]}{[1+\sfrac{\alpha}{\pi}
F(x_\pi)]} \nn \\
\tvs
&&\hspace{-26mm}\cdot\,\Big\{1-\sfrac{\alpha}{\pi}\Big[\sfrac{3}{2}\ln
\sfrac{M_\pi}{M_K}\!+\!\Delta C_1\!+\!\Delta C_2\sfrac{m_l^2}{M_\rho^2}\ln
\sfrac{M_\rho^2}{m_l^2}\!+\!\Delta C_3\sfrac{m_l^2}{M_\rho^2}\Big]\Big\}
\end{eqnarray}
where $x_P\equiv m_l/M_P$ and $\Delta C_i\equiv C_{iK}-C_{i\pi}$. The constants
$C_{iP}$ depend on the hadronic structure, and an explicit expression for the
function $F(x_P)$ can be found in ref.~\cite{ms93}.

While extracting $F_K/F_\pi$ from the ratio \eqn{RPl2}, the dominant
uncertainty is due to the CKM matrix element $V_{us}$. Thus, it is preferable
to present a value for the ratio
\begin{equation}
\label{VFPratio}
\frac{|V_{us}|}{|V_{ud}|} \frac{F_K}{F_\pi} = 0.27618\,(39)\,(27)\,(6) =
0.27618\,(48) \,.
\end{equation}
To arrive at this result, the experimental $K_{\mu2}$ and $\pi_{\mu2}$ decay
rates have been used. For the $K_{\mu2}$ decay rate, we have employed the very
recent result by the KLOE collaboration
$B[K^+\!\to\!\mu^+\nu_\mu(\gamma)]=0.6366\,(18)$ \cite{kloe05}, and all other
experimental inputs have been taken from the Review of Particle Physics
\cite{pdg04}. $\Delta C_1$ has recently been calculated in the framework of
$\chi$PT \cite{knrt00} with the result $\Delta C_1=Z/2\,\ln\,(M_K/M_\pi)$,
where the chiral coupling $Z=0.8\pm 0.2$ arises from the electromagnetic mass
difference of the pion. For the remaining constants, the generous estimates
$\Delta C_2=0\pm 1$ and $\Delta C_3=0\pm 3$ \cite{ms93} have been employed.
In eq.~\eqn{VFPratio}, the separated errors correspond to the uncertainty
resulting from the $K_{\mu2}$ branching ratio, the remaining experimental
inputs, being dominated by the $K$ lifetime, and the radiative corrections,
respectively. The total radiative correction in eq.~\eqn{RPl2} then amounts
to $-\,3.05\,(16)\,\sfrac{\alpha}{\pi}$, in agreement to the result of
ref.~\cite{fin96}, obtained in a different framework.

To extract a value for $F_K/F_\pi/F_+^{K\pi}(0)$, we now have to assume inputs
for $|V_{us}|\,F_+^{K\pi}(0)$ and $|V_{ud}|$. For both quantities, we employ
the results $|V_{us}|\,F_+^{K\pi}(0)=0.2173\,(8)$ from $K_{e3}$ decays as well
as $|V_{ud}|=0.9738\,(3)$, presented in the very recent review \cite{vus05}.
(For comparison, see however also ref.~\cite{ckm03}.) One then obtains:
\begin{equation}
\label{FKoFPoFp0}
\frac{F_K}{F_\pi\,F_+^{K\pi}(0)} = 1.2377\,(22)\,(46)\,(5) = 1.2377\,(51) \,,
\end{equation}
where the different errors correspond to the uncertainties of
eq.~\eqn{VFPratio}, $|V_{us}|\,F_+^{K\pi}(0)$, and $|V_{ud}|$, respectively.
From eq.~\eqn{FKoFPoFp0}, we observe that the uncertainty on the considered
ratio is dominated by the experimental result for $|V_{us}|\,F_+^{K\pi}(0)$.
The influence of $|V_{ud}|$ is rather small, and even increasing its error by
a factor of two practically would have no effect.

In all previous expressions the hadronic matrix elements $F_K$, $F_\pi$ and
$F_+^{K\pi}(0)$ are defined in the framework of pure QCD, i.e. in the limit
$\alpha_{\rm QED}=0$. To the quoted level of precision, electromagnetic
corrections have a crucial effect in the measured physical observables and
are taken into account through explicit correction factors, as shown in
eq.~\eqn{RPl2} for $F_K/F_\pi$ \cite{ms93,mar04}. The quoted value for
$|V_{us}|\,F_+^{K\pi}(0)$, has been derived from $K_{e3}$ data \cite{vus05},
using the electromagnetic corrections computed in ref.~\cite{cnp04}.

\section{The scalar \boldmath{$K\pi$ form factor}}

As the next step towards the determination of light quark masses, the scalar
$K\pi$ form factor $F_0^{K\pi}(t)\equiv F_0(t)$ \footnote{To simplify the
notation, from now on the superscript ``$K\pi$" on the form factors will be
dropped.} will be calculated along the lines of the dispersion theoretic
approach of refs.~\cite{jop01,jop00}, employing additional constraints both
at long and short distances. For solving the system of dispersion-integral
equations, in this approach two external input parameters are required. Like
in ref.~\cite{jop01}, these inputs will be the value of the form factor at
$t=0$, $F_0(0)=F_+(0)$, as well as its value at the Callan-Treiman point
$\DKP\equiv M_K^2-M_\pi^2$, $F_0(\DKP)$.

For $F_0(0)$, we employ an average over recent determinations from the lattice
and effective field theory approaches \cite{jop04,bec04,cir05,daw05,tsu05,oka05}
\begin{equation}
F_0(0) = 0.972(12) \,,
\end{equation}
which is also compatible with the original estimate by Leutwyler and Roos
\cite{lr84}. Together with the results of the previous section, this choice
corresponds to
\begin{equation}
|V_{us}| = 0.2236(29) \quad \mbox{and} \quad
\frac{F_K}{F_\pi} = 1.203(16) \,.
\end{equation}
The value for $|V_{us}|$ is compatible with unitarity at the $1.2\,\sigma$
level, while $F_K/F_\pi$ is about $1\,\sigma$ lower than the result of
ref.~\cite{lr84}. The ratio of decay constants is also in nice agreement with
the recent lattice results \cite{milc04}.

Because the value at the origin only concerns the global normalisation of
$F_0(t)$, from the dispersive approach a more precise determination of the
form factor shape can be obtained by first concentrating on the ratio
$F_0(t)/F_0(0)$. On the other hand, to a very good approximation, the second
required input $F_0(\DKP)$ is given by $F_K/F_\pi$, which then turns
into the relation
\begin{equation}
\label{F0CT}
\frac{F_0(\DKP)}{F_0(0)} = \frac{F_K}{F_\pi\,F_0(0)} +
\frac{\Delta_{\rm CT}}{F_0(0)} = 1.2346(53) \,,
\end{equation}
where $\Delta_{\rm CT}=-\,3\cdot 10^{-3}$ has been obtained within $\chi$PT
at order $p^4$ \cite{gl85b}, and half of this value has been added
quadratically to the uncertainty of eq.~\eqn{FKoFPoFp0}.

\begin{figure}[htb]
\begin{center}
\includegraphics[angle=0, width=8.6cm]{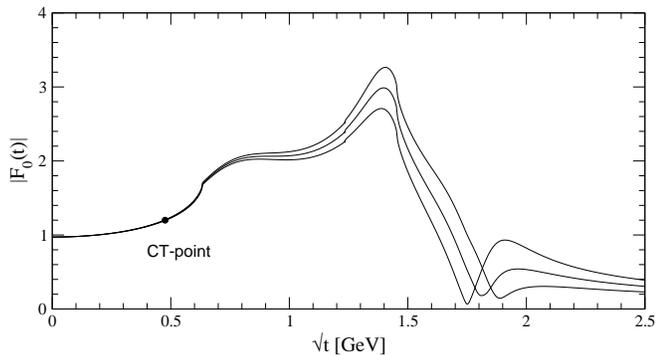}
\end{center}
\vspace{-6mm}
\caption{Result for the scalar $K\pi$ form factor $F_0(t)$ with the range
corresponding to a variation of the form factor at the Callan-Treiman-point
according to eq.~\eqn{F0CT}.\label{fig:ff}}
\end{figure}

Using this input parameter and performing an average over the different fits
to the S-wave $K\pi$ scattering data, as discussed in detail in \cite{jop01},
for the slope and the curvature of the form factor at the origin, we obtain:
\begin{equation}
\label{lambda0}
\frac{F_0'(0)}{F_0(0)}  = 0.773(21)\,\gev^{-2} \,, \, 
\frac{F_0''(0)}{F_0(0)} = 1.599(52)\,\gev^{-4} \,.
\end{equation}
Our result for the slope can also be expressed in terms of the scalar $K\pi$
squared radius, or the parameter $\lambda_0$:
\begin{eqnarray}
\langle r_{K\pi}^2\rangle &\!=\!& 6\,\frac{F_0'(0)}{F_0(0)} =
0.1806(49)\,{\rm fm}^2 \,, \\
\tvs
\lambda_0 &\!=\!& M_\pi^2\,\frac{F_0'(0)}{F_0(0)} = 0.0147(4) \,.
\end{eqnarray}
The value for $\lambda_0$ is compatible with our previous result presented in
\cite{jop04}. It is also in good agreement with the recent experimental result
by KTeV \cite{ktev04}, but is about 3$\sigma$ lower then the corresponding
ISTRA result \cite{ist04}, where however $F_0''(0)$ was found to be compatible
with zero. In figure~\ref{fig:ff}, we display our result for the the scalar
$K\pi$ form factor $F_0(t)$ as a function of $\sqrt{t}$, while varying its
value at the Callan-Treiman-point according to eq.~\eqn{F0CT}.

\section{\boldmath{$m_s$} from the scalar channel}

The numerical analysis of the scalar channel proceeds in complete analogy to
the previous analyses \cite{jop01b,jam97,jm94} where the detailed theoretical
expressions can be found. The central equation for the extraction of the light
quark masses from the scalar and pseudoscalar spectral functions takes the form
\begin{equation}
\label{sr}
u\,\wh\Psi_{th}(u) = \int\limits_0^{s_0}\! \rho_{ph}(s)\,e^{-s/u}\,ds +
\int\limits_{s_0}^\infty\! \rho_{th}(s)\,e^{-s/u}\,ds \,,
\end{equation}
in which $\wh\Psi_{th}(u)$ denotes the Borel-transformed theoretical 2-point
correlator, providing an exponential suppression of higher-energy contributions
where no experimental information is available, and $u$ is the so-called Borel
variable. Expressions for $\wh\Psi_{th}(u)$ up to ${\cal O}(\as^3)$ can be
found in \cite{jop01b,jm94} and we have also included the very recent result on
the perturbative ${\cal O}(\as^4)$ contribution \cite{ck05,bck05}.

The OPE which is employed for calculating $\wh\Psi_{th}(u)$ is valid only at
sufficiently large $u\gg\Lambda_{\rm QCD}^2$. As is well known, for the scalar
and pseudoscalar channels a breakdown of the OPE is expected to occur at a
relatively large $u$ around $1\,\gev^2$, due to the presence of non-perturbative
vacuum effects which go beyond the local condensate expansion
\cite{nsvz79,nsvz81}. Models of the correlation function based on instanton
configurations, such as the instanton liquid model (ILM)
\cite{sv93,deks96,ess98}, allow to penetrate into the region of smaller $u$.
However, as realised e.g. in \cite{mk01}, at sufficiently large
$u\approx 2\,\gev^2$, the ILM correction turns out to be rather small, and
hence we shall avoid it by choosing $2\,\gev^2$ as a lower limit for the Borel
variable, which furthermore also reduces the uncertainty from higher order
$\as$ corrections.

The relation between the phenomenological spectral function and the
strangeness-changing scalar form factor is given by
\begin{equation}
\label{rhoS}
\rho_{ph}(s) = \frac{3\Delta_{K\pi}^2}{32\pi^2}\,\Big[ \sigma_{K\pi}(s)
|F_0^{K\pi}(s)|^2 + \sigma_{K\eta'}(s)|F_0^{K\eta'}(s)|^2 \Big] \,,
\end{equation}
where $\sigma_{KP}(s)$ is the two-particle phase space factor and like in
ref.~\cite{jop01b}, we have also included the $K\eta'$ contribution
$F_0^{K\eta'}(s)$. A possible source of systematic uncertainty may be the
neglect of more than 2-particle final states, on which we comment further
below. Above the energy $s_0$, the spectral function is again approximated
by the theoretical expression $\rho_{th}(s)$.

Performing the $m_s$ analysis on the basis of eqs.~\eqn{sr} and \eqn{rhoS},
for the running mass in the $\overline{\rm MS}$ scheme we find
\begin{equation}
\label{msS}
m_s(2\,\gev) = 87.6^{\,+\,8.8}_{\,-\,6.8}\;\mev \,,
\end{equation}
at a central value $s_0=4.4\,\gev^2$, where $m_s$ is most stable with respect
to variations of $u$ in the range $2\,\gev^2<u<4\,\gev^2$. This and all other
input parameters whose variation induces a shift in $m_s$ of more than
$1\,\mev$ have been collected in table~\ref{tab1}.

\begin{table}[htb]
\renewcommand{\arraystretch}{1.2}
\begin{center}
\begin{tabular}{ccc}
\hline
Parameter & Value & $\Delta m_s$ [MeV] \\
\hline
$F_0(\DKP)/F_0(0)$ & $1.2346(53)$ & ${}^{+\,7.0}_{-\,5.3}$ \\
$F_0(0)$ & $0.972(12)$ & ${}^{+\,1.0}_{-\,1.1}$ \\
$\alpha_s(M_Z)$ & $0.119(2)$ & ${}^{-3.1}_{+3.8}$ \\
${\cal O}(\as^4)$ & ${}^{{\rm no}\;{\cal O}(\as^4)}_{2\times{\cal O}(\as^4)}$ &
${}^{+1.8}_{-1.1}$ \\
$s_0$ & $3.9 - 5.5\;\gev^2$ & ${}^{+3.2}_{-2.6}$ \\
\hline
\end{tabular}
\end{center}
\vspace{-2mm}
\caption{Values of the main input parameters and corresponding uncertainties
for $m_s(2\,\gev)$ in the scalar channel.\label{tab1}}
\end{table}

The dominant phenomenological uncertainty on $m_s$ is due to the shape of the
form factor $F_0(t)$ while the value $F_0(0)$ only plays a minor role. On the
theoretical side the main uncertainty arises from the variation of $\as$ and
to a smaller extent from unknown higher order corrections which are estimated
by either removing or doubling the ${\cal O}(\as^4)$ correction. Finally, we
have varied the parameter $s_0$ in a rather generous range. Higher order
corrections in the OPE have been included according to \cite{jop01b,jm94}.
However, they have only small influence and a variation of the corresponding
parameters induces errors on $m_s$ of less than $1\,\mev$ in all possible
cases.

\section{\boldmath{$m_s$} from the pseudoscalar channel}

In complete analogy to the scalar channel the strange mass $m_s$ can also be
extracted from the pseudoscalar channel. The phenomenological spectral function
has been modelled along the lines of ref.~\cite{mk01}, while our analysis
parallels the one of the recent work \cite{ck05}.

From the pseudoscalar channel, at an $s_0=4.2\,\gev^2$ for which $m_s$ in the
region $2\,\gev^2<u<4\,\gev^2$ is most stable against a variation of the Borel
variable $u$, the strange quark mass is found to be
\begin{equation}
\label{msP}
m_s(2\,\gev) = 97.2^{\,+\,11.3}_{\,-\,8.0}\;\mev \,.
\end{equation}
Again, in table~\ref{tab2} the input parameters and their variations which
produce a shift of $m_s$ larger than $1\,\mev$ are compiled. The most important
parameters are the decay constants of the first two excited $K$ resonances,
the $K(1460)$ and $K(1830)$, which have been estimated in ref.~\cite{mk01}.
Since especially the decay constant of the second $K(1830)$ resonance is not
very well determined, to be conservative a generous range has been assumed.
We should note, however, that employing the central input parameters of
ref.~\cite{ck05}, their results are exactly reproduced. Furthermore, the
dependence on the QCD coupling, higher order corrections and the parameter
$s_0$ has been analysed as in the previous section.

\begin{table}[htb]
\renewcommand{\arraystretch}{1.2}
\begin{center}
\begin{tabular}{ccc}
\hline
Parameter & Value & $\Delta m_s$ [MeV] \\
\hline
$F_{K(1460)}$ & $22.0\pm 2.6\,\mev$ & ${}^{+\,5.2}_{-\,4.4}$ \\
$F_{K(1830)}$ & $10^{+6}_{-10}$ & ${}^{+\,6.9}_{-\,4.6}$ \\
$\alpha_s(M_Z)$ & $0.119(2)$ & ${}^{-4.5}_{+6.6}$ \\
${\cal O}(\as^4)$ & ${}^{{\rm no}\;{\cal O}(\as^4)}_{2\times{\cal O}(\as^4)}$ &
${}^{+1.4}_{-0.8}$ \\
$s_0$ & $3.7 - 5.5\;\gev^2$ & ${}^{+2.8}_{-1.5}$ \\
\hline
\end{tabular}
\end{center}
\vspace{-2mm}
\caption{Values of the main input parameters and corresponding uncertainties
for $m_s(2\,\gev)$ in the pseudoscalar channel.\label{tab2}}
\end{table}

\section{Conclusions}

In the previous two sections, the strange quark mass has been determined on the
basis of improved results for the strangeness-changing scalar spectral function
presented in section~3, and the resonance model \cite{mk01} in the case of the
pseudoscalar spectral function. As can be observed from eqs.~\eqn{msS} and
\eqn{msP} within the uncertainties, both results are in reasonable agreement,
and thus, we are in a position to average them. Since it is difficult to assign
a precise meaning to the theoretical uncertainties, we have decided to take
the arithmetic mean, and assigned the larger error of \eqn{msS} as the total
uncertainty, which yields our final result
\begin{equation}
\label{ms}
m_s(2\,\gev) = 92.4 \pm 8.8 \;\mev = 92 \pm 9 \;\mev \,,
\end{equation}
providing the first determination of $m_s$ at the 10\% level from non-lattice
approaches. The fact that $m_s$ from the scalar channel turns out smaller than
for the pseudoscalar channel might be attributed to the fact that only
2-particle intermediate states have been included. Nevertheless, on the basis
of large-$N_C$ arguments, the contribution from higher-multiplicity states is
expected to be suppressed, and also the uncertainties in the pseudoscalar
channel are so large that at present no significance can be attributed to this
difference. Still, we plan to investigate this question further in the future.

Making use of our final result on $m_s$ of eq.~\eqn{ms}, and two particular
quark mass ratios obtained from $\chi$PT \cite{leu96},
$R \equiv m_s/\hat m = 24.4\pm 1.5$ as well as
$Q^2 \equiv (m_s^2-\hat m^2)/(m_d^2-m_u^2) = (22.7\pm 0.8)^2$, where
$\hat m\equiv (m_u+m_d)/2$, we are also in a position to calculate the light
up and down quark masses with the result:
\begin{eqnarray}
\label{mumd}
m_u(2\,\gev) &=& 2.7 \pm 0.4 \;\mev \,,\\
m_d(2\,\gev) &=& 4.8 \pm 0.5 \;\mev \,.
\end{eqnarray}

To conclude, let us compare our determinations of $m_s$ presented in eqs.
\eqn{msS} and \eqn{msP} with other recent extractions of this quantity from
sum rules, and lattice QCD. To this end, in figure~2 the $m_s$ values obtained
in this work, as well as the recent determinations from $e^+e^-$-scattering
\cite{nar05} and hadronic $\tau$ decays \cite{gjpps04}, are displayed as the
full circles. The most recent results from lattice simulations at $N_f=2+1$
\cite{hpqcd06,cppacs05} and $N_f=2$ \cite{qcdsf06,spqcd05,alpha05} quark
flavours are shown as the full squares.

\begin{figure}[htb]
\begin{center}
\includegraphics[angle=0, width=8.6cm]{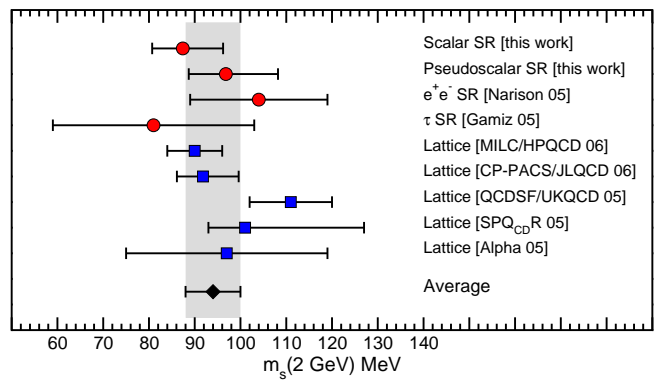}
\end{center}
\vspace{-6mm}
\caption{Compilation of recent determinations of $m_s$ from sum rules (circles)
and lattice QCD (squares).\label{fig:ms}}
\end{figure}

All results for $m_s$ are in good agreement -- perhaps with the exception of
the QCDSF/UKQCD value \cite{qcdsf06} which lies a bit high -- and thus we are
in a position to present a global average. To do this, we have calculated a
weighted average of all numbers, taking the larger uncertainty in the case
of unsymmetric errors, which leads to
\begin{equation}
\label{msav}
m_s(2\,\gev) = 94 \pm 6 \;\mev \,,
\end{equation}
where as the total uncertainty we have chosen the error of the single best
determination of $m_s$ from the lattice.

\begin{acknowledgments}
This work has been supported in part by the European Union (EU) RTN Network
EURIDICE Grant No HPRN-CT2002-00311 (M.J., J.A.O and A.P.), by MEC (Spain)
and FEDER (EU) Grants No FPA2005-02211 (M.J.), No FPA2004-03470 (J.A.O.) and
No FPA2004-00996 (A.P.), as well as Fundaci\'on  S\'eneca grant
Ref.~02975/PI/05, HadronPhysics I3 Project (EC) Contract
No~RII3-CT-2004-506078 (J.A.O) and by Generalitat Valenciana Grants
GRUPOS03/013, GV04B-594 and GV05/015 (A.P.).
\end{acknowledgments}

\bibliography{ffms06}

\begin{thebibliography}{10}

\bibitem{svz79}
M.A. Shifman, A.I. Vainshtein and V.I. Zakharov, Nucl. Phys. \textbf{B147}
  (1979) 385, 448.

\bibitem{gl85}
J.~Gasser and H.~Leutwyler, Nucl. Phys. \textbf{B250} (1985) 465.

\bibitem{nprt83}
S.~Narison, N.~Paver, E.~{de Rafael} and D.~Treleani, Nucl. Phys. \textbf{B212}
  (1983) 365.

\bibitem{jop01b}
M.~Jamin, J.A. Oller and A.~Pich, Eur. Phys. J. \textbf{C~24} (2002) 237,
  hep-ph/0110194.

\bibitem{mal99}
K.~Maltman, Phys. Lett. \textbf{B462} (1999) 195, hep-ph/9904370.

\bibitem{jam97}
M.~Jamin, Nucl. Phys. Proc. Suppl. \textbf{64} (1998) 250, hep-ph/9709484.
  Proc. of {\em QCD 97}, Montpellier, July 1997.

\bibitem{cfnp97}
P.~Colangelo, F.~{de Fazio}, G.~Nardulli and N.~Paver, Phys. Lett.
  \textbf{B408} (1997) 340, hep-ph/9704249.

\bibitem{bgm97}
T.~Bhattacharya, R.~Gupta and K.~Maltman, Phys. Rev. \textbf{D57} (1998) 5455,
  hep-ph/9703455.

\bibitem{cps96}
K.G. Chetyrkin, D.~Pirjol and K.~Schilcher, Phys. Lett. \textbf{B404} (1997)
  337, hep-ph/9612394.

\bibitem{cdps94}
K.G. Chetyrkin, C.A. Dominguez, D.~Pirjol and K.~Schilcher, Phys. Rev.
  \textbf{D51} (1995) 5090, hep-ph/9409371.

\bibitem{jm94}
M.~Jamin and M.~M{\"u}nz, Zeit. Phys. \textbf{C66} (1995) 633, hep-ph/9409335.

\bibitem{ck05}
K.G. Chetyrkin and A.~Khodjamirian, Eur. Phys. J. C \textbf{C46} (2006) 721,
  hep-ph/0512295.

\bibitem{mk01}
K.~Maltman and J.~Kambor, Phys. Rev. \textbf{D65} (2002) 074013,
  hep-ph/0108227.

\bibitem{jop00}
M.~Jamin, J.A. Oller and A.~Pich, Nucl. Phys. \textbf{B587} (2000) 331,
  hep-ph/0006045.

\bibitem{jop01}
M.~Jamin, J.A. Oller and A.~Pich, Nucl. Phys. \textbf{B622} (2002) 279,
  hep-ph/0110193.

\bibitem{pdg04}
S.~{Eidelman} et~al., {Phys. Lett.} \textbf{B592} (2004) 1,
  \url{http://pdg.lbl.gov}.

\bibitem{ms93}
W.J. Marciano and A.~Sirlin, Phys. Rev. Lett. \textbf{71} (1993) 3629.

\bibitem{kloe05}
KLOE Collaboration, F.~Ambrosino et~al., Phys. Lett. \textbf{B632} (2006) 76,
  hep-ex/0509045.

\bibitem{knrt00}
M.~Knecht, H.~Neufeld, H.~Rupertsberger and P.~Talavera, Eur. Phys. J.
  \textbf{C12} (2000) 469, hep-ph/9909284.

\bibitem{fin96}
M.~Finkemeier, Phys. Lett. \textbf{B387} (1996) 391, hep-ph/9505434.

\bibitem{vus05}
E.~Blucher et~al.  (2005), hep-ph/0512039.

\bibitem{ckm03}
{M.~Battaglia~et~al.}, CERN Yellow Report 2003-002  (2003), hep-ph/0304132.

\bibitem{mar04}
W.J. Marciano, Phys. Rev. Lett. \textbf{93} (2004) 231803, hep-ph/0402299.

\bibitem{cnp04}
V.~Cirigliano, H.~Neufeld and H.~Pichl, Eur. Phys. J. \textbf{C35} (2004) 53,
  hep-ph/0401173.

\bibitem{jop04}
M.~Jamin, J.A. Oller and A.~Pich, J. High Energy Phys. \textbf{02} (2004) 047,
  hep-ph/0401080.

\bibitem{bec04}
D.~Becirevic et~al., Nucl. Phys. \textbf{B705} (2005) 339, hep-ph/0403217.

\bibitem{cir05}
V.~Cirigliano et~al., JHEP \textbf{04} (2005) 006, hep-ph/0503108.

\bibitem{daw05}
C.~Dawson et~al., PoS \textbf{LAT2005} (2005) 337, hep-lat/0510018.

\bibitem{tsu05}
JLQCD Collaboration, N.~Tsutsui et~al., PoS \textbf{LAT2005} (2005) 357,
  hep-lat/0510068.

\bibitem{oka05}
Fermilab Lattice, MILC and HPQCD Collaboration, M.~Okamoto, PoS
  \textbf{LAT2005} (2005) 013, hep-lat/0510113.

\bibitem{lr84}
H.~Leutwyler and M.~Roos, Zeit. Phys. \textbf{C25} (1984) 91.

\bibitem{milc04}
MILC Collaboration, C.~Aubin et~al., Phys. Rev. \textbf{D70} (2004) 114501,
  hep-lat/0407028. Poster presented by R. Sugar at Lattice 06, Tucson, Arizona.

\bibitem{gl85b}
J.~Gasser and H.~Leutwyler, Nucl. Phys. \textbf{B250} (1985) 517.

\bibitem{ktev04}
KTeV Collaboration, T.~Alexopoulos et~al., Phys. Rev. \textbf{D70} (2004)
  092007, hep-ex/0406003.

\bibitem{ist04}
O.P. Yushchenko et~al., Phys. Lett. \textbf{B589} (2004) 111, hep-ex/0404030.

\bibitem{bck05}
P.A. Baikov, K.G. Chetyrkin and J.H. K{\"u}hn, Phys. Rev. Lett. \textbf{96}
  (2006) 012003, hep-ph/0511063.

\bibitem{nsvz79}
V.A. Novikov, M.A. Shifman, A.I. Vainshtein and V.I. Zakharov, Phys. Lett.
  \textbf{B86} (1979) 347.

\bibitem{nsvz81}
V.A. Novikov, M.A. Shifman, A.I. Vainshtein and V.I. Zakharov, Nucl. Phys.
  \textbf{B191} (1981) 301.

\bibitem{sv93}
E.V. Shuryak and J.J.M. Verbaarschot, Nucl. Phys. \textbf{B410} (1993) 55,
  hep-ph/9302239.

\bibitem{deks96}
A.E. Dorokhov, S.V. Esaibegian, N.I. Kochelev and N.G. Stefanis, J. Phys.
  \textbf{G23} (1997) 643, hep-th/9601086.

\bibitem{ess98}
V.~Elias, F.~Shi and T.G. Steele, J. Phys. \textbf{G24} (1998) 267.

\bibitem{leu96}
H.~Leutwyler, Phys. Lett. \textbf{B378} (1996) 313, hep-ph/9602366.

\bibitem{nar05}
S.~Narison  (2005), hep-ph/0510108.

\bibitem{gjpps04}
E.~G\'amiz et~al., Phys. Rev. Lett. \textbf{94} (2005) 011803, hep-ph/0408044.
  Nucl. Phys. Proc. Suppl. {\bf 144} (2005) 59-64, hep-ph/0411278.

\bibitem{hpqcd06}
MILC/HPQCD Collaboration, Q.~Mason et~al., Phys. Rev. \textbf{D73} (2006)
  114501, hep-ph/0511160. Poster presented by R. Sugar at Lattice 06, Tucson,
  Arizona.

\bibitem{cppacs05}
CP-PACS/JLQCD Collaboration, T.~Ishikawa et~al., PoS \textbf{LAT2005} (2006)
  057, hep-lat/0509142. Talk presented by T. Ishikawa at Lattice 06, Tucson,
  Arizona.

\bibitem{qcdsf06}
QCDSF/UKQCD Collaboration, M.~G{\"o}ckeler et~al., Phys. Rev. \textbf{D73}
  (2006) 054508, hep-lat/0601004.

\bibitem{spqcd05}
SPQCDR Collaboration, D.~Becirevic et~al., Nucl. Phys. \textbf{B734} (2006)
  138, hep-lat/0510014.

\bibitem{alpha05}
ALPHA Collaboration, M.~Della~Morte et~al., Nucl. Phys. \textbf{B729} (2005)
  117, hep-lat/0507035.

\end{thebibliography}

\end{document}